# Smart Healthcare System Implementation Challenges: A stakeholder perspective

Muhammad Hamza[1], Muhammad Azeem Akbar[1]
[1]Software Engineering, LUT University, 53851 Lappeenranta, Finland

**Abstract:**
The smart healthcare system has gained significant attention for the improvement of the customary healthcare system. The system is comprised of several key stakeholders that make the whole ecosystem successful. However, these stakeholders offer considerable challenges that need much research to address for making the system acceptable and reliable. Furthermore, very few studies examine the key challenges from the perspective of stakeholders of the smart healthcare system. The objective of this research study is to identify the key challenges associated with each stakeholder of the smart healthcare system. We have identified 27 challenges associated with eight key stakeholders of smart healthcare reported in the state-of-the-art literature. Further, a quantitative survey was conducted and the data from 85 respondents were collected in order to assess the significance of challenges in the real-world smart healthcare system. The collected data from the respondents were further analyzed using the smart-PSL (3.0). The results indicated that all the identified challenges associated with each stakeholder negatively influence the smart healthcare system.

## 1. Introduction:

Healthcare is an integral part of life. The gradual increase in the number of an aging population and associated chronic diseases such as Covid-19 pose a significant negative impact on the healthcare system [1, 2]. With the passage of time complex diseases are evolving by several factors and thus need more effective diagnosis and treatment techniques. Simultaneously, ample research on AI-aided treatment has gained significant influence over conventional treatment methods [3]. The number of aging populations will hit 1.5 billion by 2050 as predicted by the world health organization (WHO) [4]. The surge in the population and disease demands high resources of hospitals and medical practitioners that substantially can increase the treatment cost. The cost of healthcare is predicted to increase to 19.4 percent in 2027 as compared to 2017 which was 17.9 percent of gross domestic product (GDP) alone in the USA [5]. Thus, there is a need to lessen the pressure on the healthcare system by using smart healthcare while maintaining high-quality patient care, timely, efficiently, and cost-effectively [6].

A smart healthcare system has generally been recognized as the potential solution to make healthcare access and delivery easier and more cost-effective. It is comprised of various sensors, actuators, and wireless networks to get the bird eye view of heterogenous [6] and a substantial amount of research is conducted to implement the concept of smart healthcare for monitoring the patients' health with specific condition such as diabetes [7] or Parkinson's disease [8]. Smart wearable and implantable devices have widely been used to monitor the various parameters such as recording blood sugar, pulse, temperature, ECG (electrocardiogram), etc. of the patients. Apart from these, various kind of devices used for tracking and reading the ample amount of patient data. The ecosystem comprised of primarily four components i.e., end users (patients or medical practitioners), sensors that are used to collect and transmit data, the communication network over which the data transmits, and the application. The whole smart healthcare ecosystem mainly works on the three-layered architecture such as data collection layer that collects and transmit data over the network using various sensors, data storage layer for storing the collected data over the cloud, and data processing layer used for examination of patient's data by medical practitioners [9].

Medial Internet of thing (M-IoT) comprises of various stakeholder to make this smart healthcare ecosystem successful. Apart from the benefits, the whole ecosystem is still in its nascent stage and face several challenges. Several research studies have been conducted to highlight the key challenges presented in the smart healthcare ecosystem. The escalated growth in the use of smart healthcare devices automating the healthcare system with providing the potent ground to several cyber-attacks. A brief survey study conducted to highlight the potential attacks and anomalies and their detection with machine and deep learning-based intrusion detection system (IDS) [10]. Similarly, another research study highlights the future vulnerabilities presented with respect to various architectural layers [11]. A systematic study approach was conducted to represent the security challenges starting form node to the ecosystem and managing the privacy, trust and responsibility [12]. Furthermore, these smart devices send data to the third-party service provider know as cloud. The data stored on the cloud poses several technical and no-technical challenges. Yuan et al. [13] highlighted the key challenges in the area of electronic healthcare record (EHR), cloud, and big data for healthcare.

Several research studies have been conducted to identify the challenges in the domain of smart healthcare security and privacy, e-health cloud, healthcare big data [12, 14, 15]. However, smart healthcare ecosystem comprised of various stakeholders and addressing key challenges of each stakeholder will eventually help this ecosystem successful. The key stakeholders of smart healthcare system are medical practitioners, patients, operational team, regulatory bodies, manufacturer, network infrastructure provider, storage service provider, and insurance organizations [16, 17]. All these stakeholders collaboratively help to implement the smart healthcare ecosystem. However, each stakeholder individually confronting several key challenges that need to be focused. Thus, there is lake of empirical study that highlight the challenges against each stakeholder of smart healthcare system. The main contribution of this work is to identify the key challenges that are critical to focus with the intent of wide-scale adoption of the smart healthcare system.

Rest of the paper is organized as: study background and research hypothesis are discussed in section 2. Section 3 consists of research model and methodology. Similarly, Section 4 discuss the result and analysis. Section 5 covers the discussions, and the study implications and future directions are discussed in section 6. The threat to validity is presented in section 7. The conclusions are summarized in section 8.

## 2. Background and research hypothesis
### 2.1 Background
Healthcare is a crucial need of every individual. The rise in the chronic diseases such as cardiovascular, stroke diabetes, respiratory, digestive, and mental health disorders are considered the most prevalent diseases that cause high mortality in every country [18]. Most of these chronic diseases are prevalent in the rural area as compared to urban where people have less access to the healthcare facility [19]. With the passage of the time, more complex diseases are evolving thus need of more effective diagnosis and treatments. However, the birth of modern computing techniques such as Artificial intelligence, machine learning, expert systems and knowledge representation techniques altogether improved the way of diagnosis, prognosis, and so on [20]. The smart techniques incorporated with the smart IoT healthcare devices leads to the way much improvement in the healthcare sector.

Medical Internet of Things (m-IoT) or smart healthcare system is a network of objects like sensors, actuators that converse with each other in a diverse network with or without the assistance of computer. The rise in these smart devices is predicted to reach up to 2.1 trillion by the end of 2025 [21]. The IoT in healthcare sector can assist monitor the patient health remotely taking the advantage over conventional health monitoring system where a doctor can observe a patient in a specific time contrary to considering the emergency that can occur at any time. These smart portable healthcare devices such as smart wearable devices permitting patient to wear them round the clock to monitor their health and eventually reducing the pressure on hospital resources. Furthermore, the data generated by these devices and transmission of these data to medical practitioner can help better decision making. Apart from the undeniable benefits of medical internet of things (m-IoT), there are several challenges occur when using the devices when remotely monitoring such as security and privacy issues.

Smart healthcare ecosystem is still in its early stage of development thus posing several challenges that could hinder the successful adoption of IoT healthcare system. Several research studies have been conducted to identify the key challenges in the smart healthcare ecosystem. The main objective is to safeguard the whole ecosystem considering the security and privacy requirements [22]. Algarni et al. [23] identified the potential security challenge that can arise with the passage of time. The author further presented the open issues with respect to the layers of IoT. Similarly, author in [24] developed a taxonomy of potential attacks and classified into six categories considering the architecture, security requirements such as (confidentiality, integrity, authenticity), smart device functionalities, communication channel, and corrupted or modified packets.

The smart devices monitor the data and send it to the cloud for storage as the miniaturized nature of devices do not permit to store manipulated data locally in the device. Thus, storing data on the cloud offers several challenges. Sadeeq et al. in [25] explained the digital transformation of healthcare system and key challenges presented in the cloud services provider. Similarly, Ahmad et al. [26] presented a review of cloud computing security challenges related to data location, storage, confidentiality, integrity, and availability. Similar another review study has been conducted on the combination of cloud and IoT and the possible challenges [27].

The smart healthcare or medical internet of things (M-IoT) ecosystem comprised of several stakeholders such as medical practitioners, patients, operational team, manufacturer of devices, regulatory bodies, network infrastructure provider, insurance organizations. All the stakeholders will collaboratively assist this ecosystem for its successful adoption. However, each stakeholder offers serval challenges that need much attention to address that will eventually make the whole ecosystem acceptable and successful. Various research studies have been conducted to identify challenges in M-IoT but there is no single study exists which present the challenges in all stakeholders. This research studies have been conducted to identify the key challenges presented in all stakeholders of smart healthcare ecosystem. By reviewing the existing literature, we identified the challenges that are critical for each stakeholder of smart healthcare system, the brief explanation is given below.

## 2.2 Research hypothesis (stakeholders and respective challenges)

**Stakeholder No. 1-Medical Practitioner**

*Ch.1 (Lake of authentication)*
The smart healthcare system is considered as an indispensable sector which has ubiquitous impact across the globe. However, this system can be collapsed if the key challenges are not timely addressed. We have identified that the lake of authentication system for medical practitioners against the hospital can be a critical challenge [28]. There are few systems, which record the profile of medical practitioners i.e., the authentic work experience in the medical field, the core expertise, the number of patients medical practitioners have so far examined, the success and failure rate of those cases, expertise in using the modern equipment, and the novelty in his core domain. There should be a system that trace and authenticate the previous history of medical practitioners in term of their specialty. Various federated authentication system has been proposed to validate the authenticity of medical practitioners [29]. For example, Catalan digital health system [30] has proposed the authentication-based system for the doctors against hospital system using login credentials or X.509 certificate. However, these systems just validate the authenticity of medical practitioners but lacking in authentication of their previous profile. It has been noticed that various medical practitioners with obsolete experience are still performing activities in most of the hospitals. The patients are more concerned about the profile of medical practitioners. The trustworthiness of the developed system that validate the profile of medical practitioners can impose the positive influence on patients and overall IoT healthcare ecosystem. However, the missing of authentication element of medical practitioners in the system could hinder the satisfaction and treatment of patients.

> ***Conclusion (Ch.1):*** *The lack of authentication of medical practitioners' profiles poses a negative impact on the trustworthiness of smart healthcare systems.*

### Ch.2 (Lake of Smart devices training certificates)

Smart healthcare has been exerting the undeniable impact for modernizing the outdate healthcare system. The increasing number of smart devices having complex architecture improving the quality and simplifying the healthcare system. However, the heterogenous nature of the devices can be complicated to utilize appropriately. These smart devices comprise of various features that could be complex to use [31]. Further, there are significant barriers and enablers while adoption of the new technology [32, 33]. It is reported that various manufacturers are developing devices based on their own developed architecture. The lake of standardized architecture that is acceptable universally may pose several challenges [34]. However, complex nature of smart devices can be complicated to use and thus need a proper training [28]. The courses required to be the medical practitioners lacking in the including the role of smart devices in the modernization of the healthcare system. Furthermore, professional medical practitioners are still using the obsolete healthcare devices and are not familiar with smart devices [32, 33]. Thus, the implementation of smart healthcare system cannot be effective until the medical professionals are not educated about the significance and use of smart devices. The manufacturer of smart devices should introduce the training certificates. The Governmental organizations and hospitals must require the medical practitioner to have pertinent certification before utilizing the smart devices.

> ***Conclusion (Ch.2):*** *The lack smart devices training certification for medical practitioners could negatively influence on the implementation of smart healthcare system.*

> ****Hypothesis-1 (H1):*** To summarize it is hypothesized that Ch.1 and Ch.2 negatively impact the Medical Practitioner and need consideration for the successful implementation of smart healthcare.

**Stakeholder No. 2: Development (manufacturer)**

### Ch.3 (The lack of interoperability among the heterogeneous platforms and standards)

Smart healthcare devices comprise of various sensors that collects and transfer data to other devices in an interconnected network. The significant increase in the development of these devices demands the interoperability among different entities. These devices are developed by different manufacturer, and each has its own proprietary protocols, devices, API, and data formats that raise the concern "The Internet of Things Might Never Speak a Common Language" [35, 36]. The interoperability becomes significantly challenging when it comes to smart healthcare as it connects the devices and people accompanying devices [37]. The heterogenous nature of these devices could enforce significant issue pertaining to security, privacy, and communication [38]. These devices could open concerns related to connecting medical devices using wireless network technologies. Although Various standards have been developed to accomplish the interoperability highlighted in [39] but lacking in the development of universally accepted standards for interoperability and security among the heterogeneous platforms and standards. IoT organizations should develop a standardized model to ensure the smart medical devices operate properly when connected using various type of wireless communication technologies making it truly interoperable.

> ***Conclusion (Ch.3):*** *The lack of interoperability among the heterogeneous nature of smart medical devices could negatively influence the smart healthcare system.*

### Ch.4 (Lake of self-learning and self-improvement methods for devices)

The rising aging papulation all over the world significantly require the skilled healthcare worker for their care and treatment [40]. Thus, lake of availability of skilled medical staff could significantly increase the mortality rate especially in epidemic i.e., Covid-19. Several smart medical devices such as glucose monitoring, Parkinson's disease monitoring, heart-rate monitoring is developed to monitor the different health aspects of patients. Furthermore, these devices have enabled remote patient monitoring (RPM). However, these devices lacking in recommending the rehabilitate treatment in emergency cases. These devices monitor, collect, and share the data to medical staff for further recommendation of treatment thus causing delay and burden on the hospitals. Therefore, precisely and well-timed treatment can only be possible on the fast patient evaluation. Few studies have highlighted the importance of machine learning based recommendation treatment for breast cancers [41]. However, these smart wearable devices lacking in the development of rehabilitant methods for the further evaluations of collected data. These smart devices should be equipped with self-learning techniques such as artificial neural network (ANN), genetic algorithms (GA), ant colony optimization (ACO) and simulated annealing (SA) that could be used for data analysis and recommendation of the optimal solution [42]. The self-recommendation treatment by smart devices could help in fast recovery for patients and lessen the burden on the hospitals.

> ***Conclusion (Ch.4):*** *The lake of self-recommendation and improvement of smart medical devices poses negative influence on the smart healthcare system.*

### Ch.5 (Lake of ethics considerations)

The smart healthcare devices are growing with huge popularity among people. This rise also resulted in perceiving several IoT system failures, which also embrace the ethical issues associated in these devices [43]. These smart devices collect unnecessary information from users without their consent and inform to others. Currently, the contact tracing app has widely been used for reducing the damage of pandemic or epidemic such as Covid-19. These apps monitor or trace the user contacts without their permission rising the privacy disclosure concern about (49%) who use these apps reported by [44]. The rise in the ethical issues with the fast-growing IoT phenomena has encouraged companies, governmental, and standardized organizations to devise the ethical code of conducts and its real implementation. There are various ethical guidelines devised by different organization by considering the ethical issues associated with smart devices [45]. However, these key principles are lacking in actionability while development of smart devices. There is need to develop a framework that could highlight the key ethical principles and the implementation of those principle in the development process.

> ***Conclusion (Ch.5):*** *The lake of ethically aligned smart medical devices could negatively impact on the smart healthcare system.*

### Ch.6 (Size limitation of smart devices)

Smart healthcare devices are microscopic in size to be implanted into the patient body to monitor the health status. These miniaturize devices typically consists of low computational power, limited memory 64KB to 640KB, and battery power, performance cost that result in need of frequent recharging. These limitations could be challenging for manufacturer and software developer to implement advanced security algorithms. The breach of the security of these smart devices could lead to vary dangerous situations for patient.

However, narrow research has been conducted on the development of lightweight cryptographic algorithms surveyed in [46]. There needs much research to develop resource constrained cryptographic algorithms for smart healthcare devices[47, 48]. Furthermore, these devices must be equipped with longer battery timing from prevention of network failure [48, 49].

> ***Conclusion (Ch.6):*** *The size limitation of the smart devices limits various features to be added for making smart healthcare ecosystem successful.*

### Ch.7 (Lake of software updates)
Smart healthcare devices are come with built in features for patient's health monitoring for specific disease i.e., diabetic mentoring system. However, the irrefutable advancement in the software industry and diagnostic techniques with the passage of time motivate frequent updates for smart devices [50]. Furthermore, these devices are prone to several security and privacy attacks. The implantable smart healthcare devices must have a mechanism to update its firmware when the latest techniques to identify and resolve the disease are introduced [51]. These updates can further secure the device form potential security attacks. The firmware that could be updated can be beneficial in case of pandemic. For example, the current covid pandemic has been spreading rapidly. Thus, if these smart devices subject to update based on the latest techniques developed for detecting the covid could reduce the impact of pandemic over the globe. Every person in near future be wearing the smart devices thus need to develop a firmware that subject to accept the update to reduce the latest pandemic effect.

> ***Conclusion (Ch.7):*** *The lake of timely software update in the smart devices could pose negative impact on the smart healthcare system.*

### Ch.8 (Lake of robust framework for scaling IoT infrastructure)
The use of smart devices is increasing tremendously. It is predicted that these devices will grow up to 80 billion to be connected to the network by the end of 2025. The logic is developed, deployed, and executed directly on the devices, thus changed the way of designing, developing, deploying, and managing these smart devices. These devices need to be scalable to meet the changing demand with innovative approaches. The heterogenous nature of these devices is a massive challenge to deal with the scalability of IoT devices. Several techniques have been introduced to address scalability issues such as automated bootstrapping [52], Controlling IoT data pipeline [53], microservice architecture[54], and many other. However, several research challenges are still need much attention to make IoT acceptable. For example, lake of standardized protocol that could accommodate and identify newly installed device in the network. Similarly, Lake of standardized framework for authentication and authorization of IoT devices.

> ***Conclusion (Ch.8):*** *The lack of framework to manage the scalable IoT infrastructure could pose negative impact in future for smart healthcare system.*

> ***\*Hypothesis-2 (H2):*** The manufacturer of smart healthcare devices needs comprehensive research study over handling the key challenges [Ch3-Ch8] that could negatively influence on the smart healthcare ecosystem.

**Stakeholder No. 3: Big data management**

### Ch.9 (Data volume and velocity issues)
Smart medical devices are constantly generating huge amount data that is surpassing the current market to generate the value from it. The transfer of health data to the cloud has been increasing exponentially from

1000 petabyte to 20ZBs. Hospitals are considering integrating the analytical tools i.e., Hadoop, hadapt, cloudera for the management of five V's Volume, Velocity, Variety, Veracity and Value of data [55, 56]. However, there still exists so many limitations that researchers need to address. The lake of common data warehouse which can be used to store the data originating in different formats from different smart devices. The collected data can be unstructured or in non-standardized form that could compromise the overall quality. Wearable or implantable smart devices having limited battery and computational power cannot use the advanced data analytical frameworks on it leading to generate the huge amount of data for storage provider. The analysis of huge amount data and generating the useful report is still challenging in case of pandemic such as Covid-19.

> ***Conclusion (Ch.9):*** *The lake of common data storage warehouse for different data format poses negative influence on smart healthcare system.*

### Ch.10 (Heterogenous data format)
Digital healthcare devices that collect data originated from wearable or implantable devices admires various challenges pertaining to data management. These devices are different in working for different body parts and collect different format of data. The range, nature, volume, and rate are different from different smart medical devices [57]. For an instance, Electro Cardio Graphic (ECG) data are recorded in XML format and respectively other formats. The collection and analysis of heterogenous data poses serious data management issues thus delaying in the patient health evaluation. There are various domains where practitioners worked to achieve the data protocols[58] but there is a lake of framework that collect data from all wearable or implantable devices and then categorizes the collected data to make the single report. On the doctors end all the devices must be reporting the collected data in a single interface for analyzing the overall body status. This could help doctor better prescribe the medicine. In the obsolete practices, the patient just describes the one disease and doctor prescribe the medicine pertaining to that disease. This practice poses several side effects on the patient health. On the other hand, if the system shows to doctor the health status of all body parts, this could help doctor to understand the overall health of patient and would assist medical practitioners to prescribe the medicine that do not have side effects on other body parts.

> ***Conclusion (Ch.10):*** *The lake of framework for managing heterogenous nature of data generated by smart devices could negatively impact the whole healthcare system.*

### Ch.11 (Lake of techniques for extraction of useful data from the patient records)
Healthcare devices continuously monitor and collect the data of patient's health. The collected data is further stored onto some storage provider. Smart devices will be sending data to storage provider. The abundance of data may be difficult to analyze. However, several tools are developed by researchers, pharmaceutical companies, and health-care providers [59] for the extraction of useful data form the stored patient's data. It is still challenging for the collection of clean, formatted, thorough, and precise data that would help medical practitioners to prescribe the best possible medical treatment.

> ***Conclusion (Ch.11):*** *The lake of patient's data analysis methods has negative impact on smart healthcare system.*

### Ch.12 (Lake of data integrity)
Data generated by the healthcare smart device are so crucial for medical practitioners to take a decision and long-term strategic planning. Inaccurate generated data to be submitted to the medical practitioners can put patient health in danger. The healthcare data comes from various sources such as clinics, medical institutions, laboratories, and hospitals need a robust authentication system to ensure the data authenticity [59]. The healthcare matrix and definitions are repetitively altering and there is a lake of discrepancy in healthcare definitions that could lead to incorrect record of data. Furthermore, Medical laboratories, Pharmacies should be register with the regulatory bodies and only registered laboratories, pharmacies can deal with the patients. Integrity should be maintained during the document control. Few research has been conducted on ensuring the integrity for managing the EHR record [60]. The integrity and authentication are main concerns and can be life threating if compromised by changing of received data.

> *Conclusion (Ch.12):* The lake of data integrity could negatively impact on the healthcare system.

> ***Hypothesis-3 (H3):*** The smart devices will be generating huge amount of data and eventually be raising [Ch9 – Ch12] challenges that could pose negative impact on the whole ecosystem.

**Stakeholder no 4: Patient**
*Ch. 13 (Lack of trust)*
The concept of smart healthcare is promptly gaining popularity among peoples and healthcare practitioners. Apart from the rise, the smart healthcare has manifold concerns that could relegate the reputation if timely not addressed. The trust is considered the main challenge because these smart devices are intimately related with people and medical practitioners. Most of the functionalities are provided by the third-party vendors and lake of authentication of third-party vendors could be challenging. Furthermore, patients are more concerned about their privacy of personal information, location, transparency, and accountability. Few studies have been conducted to address the trust factor in IoT [61, 62]. However, there is still huge gap for research to identify and address the key factor compromising the trust in IoT. Similarly, the emotion detection devices are being used in the healthcare sector to stimulate patients' emotions. The devices further used for facial recognition without user consent raising the trust issue.

> *Conclusion (Ch.13):* The lake of trust on the smart devices may negatively impact the smart healthcare system.

**Ch. 14 (Illiterate patients)**
Smart medical devices comprised of several smart features. The smart wearable device can measure the blood pressure, heart rate, accelerometer, pedometer, and other activities. However, these devices use medical terms such as systolic and diastolic that are difficult for the illiterate patient to understand. The lake of ability to understand, learn and evaluate the information shown on the smart devices could be challenging [62]. The illiteracy rate in the developing countries are more than the developed countries [63]. Furthermore, every patient may not understand medically used terms and scales for measurements of different body part's health status [64]. The lake of understanding of medical terms can be challenging for the patient health in case there is no doctor for patients monitoring [64]. Therefore, smart devices provider should offer a short course-based training in form of video in patient's native language on how to use these devices. Furthermore, the device should be intelligent enough to understand and notify to rescue service if in case critical situation of patient. This would help manage the pandemic effectively. The lake of knowledge and early detection of the disease effected many countries due to recent outbreak of Covid-19.

> *Conclusion (Ch.14):* Patient illiteracy towards the use of smart medical devices could negatively impact on the healthcare system

> ***Hypothesis-4 (H4):*** Patients are the end users of smart devices and could face [Ch13-Ch14] that need immediate resolution for the successful implantation of smart healthcare phenomena.

**Stakeholder no 5: Security and Privacy**
*Ch.15 (Lake of layer-based security mechanism)*

The escalated growth in the smart healthcare devices has been improving and reshaping the healthcare system. Alongside these advantages, the explosion use of these devices have paved the potent road to various cyber-physical attacks [65]. These smart devices comprise of RFID, wireless sensor network, cloud computing, embedded system, and the architecture that enable working of smart devices [66]. There are miscellaneous three, four, five, and seven-layer architecture, which are accepted by various professionals to have a visual sculpture of this technology and the lake of one standardized architecture make this phenomenon more prone to security attacks [67, 68]. These layers are perception, network, support, and application layer. mHealth consists of mainly three-layer architecture such as perception, data storage, and data processing layer. Each layer is prone to significant threat by the intruder. Perception layer can be subjugated to node capture and cloning, Eavesdropping, Jamming attack, resource depletion attack, relay attack. Similarly, network layer can be exploited by man-in-the-middle, routing attack, DDoS attack, sybil attacks. Support layer is prone to DoS and malicious insider attack. The application layer could be compromised by Phishing attacks, Malicious node injection, Session Hijacking attacks. Various research studies have been conducted to prevent from several attacks but lacking in developing the method to prevent from all attack on each layer [10]. Machine learning based frameworks are developed for detecting malicious activity in the hospitals, but these frameworks cannot be used in wearable and implantable devices [69]. The lake of one standardized architecture and lightweight cryptographic are another challenge for security experts to deal with. The powerful search engine such as shodan which can be used to locate the internet connected device make vulnerable to the smart wearable devices to several attacks.

> ***Conclusion (Ch.15):*** *The lake of layer-based security mechanism poses negative impact on smart healthcare system.*

*Ch.16 (Lack of intelligent vulnerability assessment technique)*

Smart devices are miniaturized in size having low computational power and memory size making it difficult for implementing cryptographic algorithms. Further, the heterogeneous nature of these devices impeding researcher to develop the automated detection and recovery of vulnerabilities [70, 71]. Lake of availability of real time datasets for detection and timely mitigation of security vulnerabilities is challenging for researcher. Similarly, constant change in the functionality of the network and lake of single automated binary patch posing serious challenge to the researcher. The database that records and maintain the structured information of exploits and vulnerabilities is still in its nascent stage.

> ***Conclusion (Ch.16):*** *The lake smart device vulnerability assessment methods pose negative impact on the healthcare system.*

*Ch. 17 (Lake of framework for detecting malicious IOT devices)*
The immense surge in the smart healthcare devices benefiting the society day by day. The rise in the number of devices posing serious challenge to the security experts as these devices can be used for malicious perspective [72]. The heterogenous nature of the IoT environment and the competition among technology giant developing updated devices without considering the crucial security perspective. The intruders on the other hand have been taking advantage of this loophole. IoT smart devices can be used by attacker for malicious purpose such as sending false information to other devices. Various research studies have been conducted to detect the unusual behavior of traffic in IoT healthcare. For example, [73] developed a framework for detecting the malicious activity in the IoT healthcare. Similarly,[74] developed a lightweight cryptographic framework for detecting the traffic at edge gateway. However, a little research has been conducted to identify the malicious device in a network. The malicious device can be used over and over again and if these devices are not registered with the user personal credentials, then identification of the attacker could be challenging. Thus, smart medical devices need to be registered with users' personal information in case to catch the intruder.

> ***Conclusion (Ch.17):*** *lake of malicious devices detection method poses negative influence on the smart healthcare system.*

> ****Hypothesis-5 (H5):*** Smart healthcare devices are more prone to security and privacy threats and could face [Ch15 – Ch17] will eventually be collapsing the system.*

**Stakeholder no 6: Network Infrastructure**

*Ch. 18 (Security in IoT cloud)*

The exponential growth of IoT devices lead to produce abundance amount of data. The data generated or collected by these devices cannot be stored locally as these devices are equipped with low computational and memory size thus motive to outsource the data. The concept of storing the data to cloud expanded quickly in recent years [75]. Outsourcing of the data to cloud poses several security threats such as cloud pooling. Privacy and confidentiality of patient's medical record is also an important concern as patient do not want to share the sensitive information of his health such as cancer and HIV reports. The privacy concerns arise when the record of patient is shared among third party services provider [76]. Similarly, lake of security, misconfigured devices and network could breach the privacy and confidentiality of patient's data. Furthermore, the cloud services provider stores the data originated by the smart device and lacking to ensure that received data is not modified intentionally or accidentally.

***Hypothesis-18 (H18):*** *Security concerns in the IoT cloud poses negative impact on smart healthcare system.*

> ***Conclusion (Ch.18):*** *Security concerns in the IoT cloud poses negative impact on smart healthcare system.*

*Ch.19 (Client Management issue)*

The surge in the use of smart devices and the data these devices outsource to cloud has posed client management issues. Client management issue has different aspects such as client experience which ensure to provide the best customer services to their users. The patients are more concerned about the cloud service provider as there are several service providers such google, Microsoft, Yandex, and choosing a verifiable company is still difficult for the patient. The lake of trust on the cloud service provider could be critical as patient will be storing his personal health information [77, 78].

> ***Conclusion (Ch.19):*** *The lake of user management framework by cloud service provider could negatively influence the smart healthcare system.*

*Ch.20 (Lake of processing information of the cloud provider)*

Smart devices collect and send patient sensitive data to storage service provider for the examination by the medical practitioners. The huge amount of raw data of patients stored on the cloud could be difficult for medical practitioners to examine. However, this huge amount of data needs to be processed to obtain the knowledgeable information. The processing of patient sensitive data by third party services provider could breach the confidentiality [79]. On the other hand, a single medical practitioner or team of doctors need to access the patient data simultaneously thus lake of authentication of medical practitioners who want to access the data is a challenge. Furthermore, incomplete access and unsuccessful communication between care team creates medical errors.

> ***Conclusion (Ch.20):*** *The lack of data processing by the cloud could pose a negative influence on the smart healthcare system.*

### *Ch.21 (Lake of reliable connectivity)*

The number of patients across the world is rapidly increasing as compared to the doctors. The lake of availability of professional medical practitioners leads the health care organizations to monitor patient's health remotely. Smart devices send the data to cloud and then medical practitioner can access the record of patient to examine. Further, the online telemedicine system helped to provide effectively and timely medical services [80]. However, these smart devices need stable internet connection for continuous data sharing and monitoring. Patient with critical condition need to be stayed on the internet connected areas. A small delay in the internet connection can turn critical situation for the patient [81]. The concept of telemedicine and monitoring of the data from cloud could be difficult in the countries where there is not stable connect. Thus, the concept of smart healthcare could be limited to the developed countries and the underdeveloped countries would be deprived of the benefits of the smart healthcare.

> ***Conclusion (Ch.21):*** *The lake of stable internet connection could negatively impact on the implementation of smart healthcare system.*

### *Ch.22 (Lack of e-Health Cloud design and development standards)*

The smart healthcare devices continuously send data to the e-health cloud services provider. The availability of data is crucial for the medical practitioners for better observation of patient's health. However, Cloud services could experience failure due to several attacks and availability could compromised in cloud as compared to organizational internal infrastructure [82]. Apart from this, multiple cloud service provider can provide different services. For example, one service provider can provide storage and processing services for HD medical images and other services provider can provide the data mining and analysis services. Thus, the lake of interoperability is a key challenge for the services providers [83]. The hospitals will consider e-healthcare cloud for implementing the smart healthcare ecosystem, thus requires significance amendments in their existing clinical and business process. A defined set of protocols are required to implement the smart healthcare ecosystem and to make compatible with the existing system working in the hospitals. There are few standards and classifications developed for health information systems such as International Classification of Diseases tenth revision (ICD-10) issued by the World Health Organization (WHO) and The Systematized Nomenclature of Medicine (SNOMED) [84]. However, the lake of legislation standards for inter-operability, medical informatics, policies, and transmission methods in e-Health Cloud still require adequate research.

> ***Conclusion (Ch.22):*** *The lake of legislation standards for inter-operability, medical informatics, policies, and transmission methods in e-Health Cloud.*

> ***\*Hypothesis-6 (H6):*** Network infrastructure is the key stakeholder for implementing smart healthcare system. However, there need huge research to address [Ch.18 – Ch.22] that could negatively influence on the system.

### *Stakeholder no 7: Regulatory Bodies*
### *Ch.23 (Lake of stakeholder's collaborations)*

The developed countries are more focusing on the developing smart healthcare ecosystem for early intervention and prevention form the disease. However, there are several factors that need much intention to ensure the successful implementation of this ecosystem [28]. The lake of partnership between public and private stakeholder hindering successful implementation of smart healthcare. Collaborative working of the stakeholder such as, Government agencies, technology companies, healthcare and life sciences player, media and NGO/NPO, social care entities, and patients can make this ecosystem successful. Patients can collaborate by sharing their experience and behavior towards the use of the smart devices. Similarly, technologies companies should work collaboratively to better handle the key challenges. Thus, there is need a central collaboration among all stakeholders to organize plan for the development of acceptable smart healthcare ecosystem.

> *Conclusion (Ch.23):* Lake of stakeholder's collaboration hider the successful implementation of smart healthcare system.

*Ch. 24 (Lake of global consensus regulations)*

The smart devices with continuously evolving features and methods are coming into the market. These devices such as wearable and implantable devices used in the patient's body for monitoring the health status need to be approved form the regulatory bodies such as food and drug association (FDA), Health Insurance Portability and Accountability Act (HIPPA) [85]. However, the regulatory bodies and legal requirements that govern the healthcare system are sometimes changing in their existing rules and regulation creating a market barrier for the companies developing smart healthcare devices. Furthermore, nearly all developed states have formulated their own individual regulatory standards on the manufacturing and use of smart healthcare devices. The manufacturer developing smart devices have to compliance with these standards i.e., Apple watch series is approved by FDA and CE marked. Thus, smart devices need to qualify based on the definition of applicable regulations and nearly every state has its own developed Medical Device Regulation. The devices developed in China need to compliance with the China's MDR as well as European MDR in case to use these devices in European countries. Thus, the lake of globally developed MDR creating barriers for the manufacturers to develop smart devices to be used in all states. This could create big challenge is case of pandemic such as Covid-19. The device developed by China for the fastest prediction of Covid in the patient need to compliance with the European of US MDR in order to be used in these countries.

> *Conclusion (Ch.24):* The lake of globally recognized consensus for a manufacturer could negatively impact on the healthcare system.

> ***Hypothesis-7 (H7):*** Regulatory bodies decisions are most important for the success of smart devices manufacture. Thus [Ch.23 – Ch.24] are critical challenge that could have eternal impact.

## Stakeholder no 8: Operational Team
*Ch.25 (Lake of smart healthcare infrastructure management professionals)*

Healthcare devices is one of the fastest growing ecosystems in IoT market as predicted to reach $176 billion by 2026. These smart devices offer several opportunities for medical practitioners as well as patients and being used in hospitals specially in intense care unit (ICU). The complex nature of these devices and heterogenous network infrastructure need professional IT team to manage this ecosystem [86]. However, there is substantial lake of qualified IT specialists that can help in implementation and evaluation of these smart devices with deep understanding of the healthcare system and the patient. A little negligence managing these devices could pose serious disruptive impact on the patient health. Further, the lake of highly qualified IT experts in the medical field can result in poor purchasing of these smart devices. Thus, there is need to train the operational team members before implementing and evaluating these devices.

> *Conclusion (Ch.25):* The lake of professional for managing smart healthcare system could fail adoption of healthcare system.

*Ch.26 (Workflow disruption during implementation of smart healthcare system)*

Smart healthcare ecosystem is widely accepted and implemented in various cities of the developed countries. The appropriate implementation of novel system demands IT experts. Thus, the unavailability of on-board IT experts requires many hospitals to outsource the task of implementation of smart healthcare ecosystem. The outsourced IT companies could get exclusive access to the patient data, software development, and maintenance. The access to the whole ecosystem by third-party services provider in the hospital could breach the confidentiality of data. Another challenge that needs to be considered is the workflow disruption during IoT healthcare ecosystem implementation. First implementation of new ecosystem can impose substantial disruption in usual clinical and administrative workflows.

> ***Conclusion (Ch.26):*** *The lack of trained workflow for smart healthcare device implementation could negatively impact.*

*Ch.27 (Lake of cyber security management professionals)*

The smart health-care ecosystem composed of several smart wearable and implantable devices that continuously monitor the patients' health in home or in Intensive care unit (ICU). However, the ecosystem is still in its nascent stage and prone to various security and privacy threats. The patients in intensive care unit (ICU) need continuous monitoring of their health. The hospital IoT network are more prone to DDoS attacks and several attacks that need considerable capital and efforts in protecting the systems. The frequency of cyber-attacks in the healthcare industry has been increasing [87]. Further, disruption in the network due to these kinds of attacks could cause patient's death. Thus, there is need of cyber security management professional in the hospital that can timely reinstate the ecosystem if attack occur. The cyber security team would continuously be monitoring the network and could response instantly in case of attack occurrence.

> ***Conclusion (Ch.27):*** *The lack of cyber security management professionals in the hospital could down the healthcare system.*

> *\* Hypothesis-8 (H8): Implementation of smart devices and networks in the hospital requires an experienced operation team. Thus, a need for planning is required to address the [Ch.25 -Ch27] challenges that could impose a negative impact on the ecosystem.*

## 3. Research model and Methodology

The aim of this research work is to studies the key stakeholders of smart healthcare system and the challenges they are facing. To come up with study objective, a focused literature review was performed, and enlist the numbers of challenges reported by researchers against stakeholders of smart health care system. The identified list of challenging factors was discussed with research team and research advisor. Based on the discussion, the challenging factors were finalized against each stakeholder. Furthermore, hypothesizes were developed based on the literature discussion. In next step, we develop the research model (Figure 2), using the proposed hypothesizes. In order to check the implacability of proposed hypothesis in real-world industry, we performed the questionnaire survey study with industry experts. The steps adopted for this research work is give in Figure 1 and discussed in the sub-sequent sections.

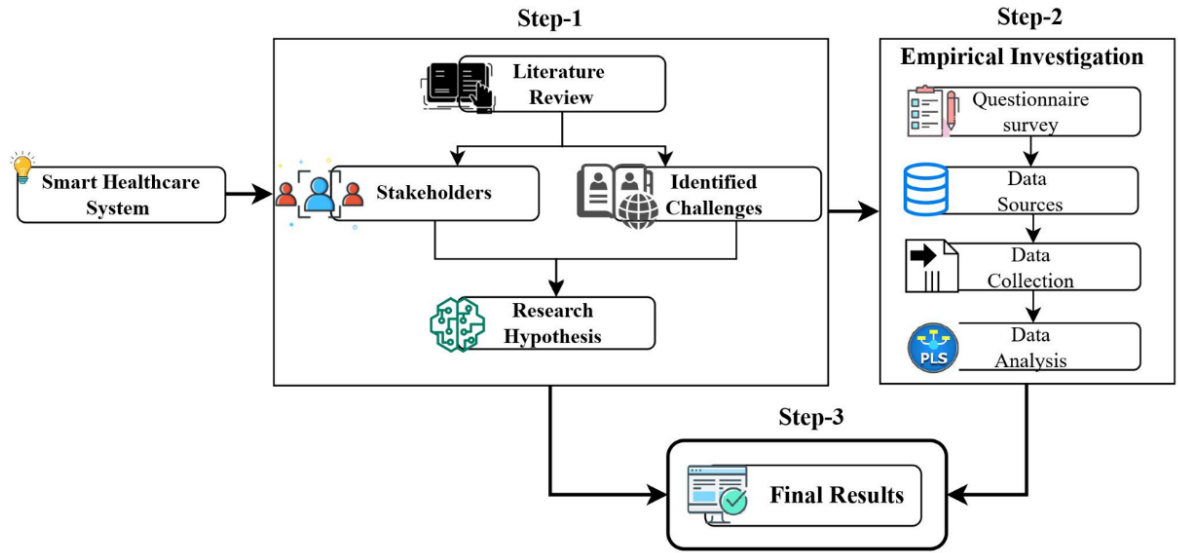

**Figure 1:** Used research process

### 3.1 Theoretical bases for the proposed framework
#### 3.1.1 Actor network and stakeholder theory

The stakeholder theory and Actor network theory is used to devise the theoretical framework for this research study. The stakeholder theory is used to identify, mapping the stakeholders, and who and what really influence in the accomplishment of the project. The stakeholder theory is limited to customers, end-users, and project sponsors but offers a model of cooperation. However, the proposed research study comprised of human and non-human stakeholders which are significant for the accomplishment of the smart healthcare concept. To address the concern, actor network theory is implemented.

Actor network theory is adopted to embrace not only people but objects and organizations. The main focus of the ANT is to emphasis on the inanimate objects and their influence on the actual process. It provides the sociotechnical perspective which enables to analyze the complex interaction between human and no human factors [112]. In ANT, there are certain set of theoretical tools and vocabulary that describes and theories the concept of influencing stakeholders in smart healthcare system.

In ANT, an actor is described as "any element which bends space around itself, makes other elements dependent upon itself and translates their will into the language of its own". Actor can be social, technical objects, and artifacts [112] and treated as equally important in devising the network. we identified eight stakeholders referred as actors in the smart healthcare system concept: medical practitioners, development (manufacturer), big data management, patient, security and privacy, network infrastructure, operational team, regulatory body.

### 3.2 Research Model
The aim of anticipated research model is to examine the relationship between the identified key challenges corresponding to each stakeholder of smart healthcare and its successful adoption in real-world. We further analyzed the significance of the identified challenges that are hindering the successful implementation of smart healthcare system. The proposed model is developed based on the existing state-of-the-art literature of smart healthcare system (Figure 2). The developed theoretical model is used for the assessment of the relationship among the dependent variable i.e., smart healthcare system and the independent variable i.e., proposed hypothesis. The proposed hypotheses are comprised of twenty-seven independent variables i.e., Ch1-Ch27. The multiple linear regression equation of the proposed is as follows:

(1) Smart healthcare system = α0+α1y1+α2y2+α3y3+α4y4+α5y5+α6y6+α7y7+α8y8

where α0, α1, α2, α3…. α8 are coefficients and y1 to y8 are independent variables.

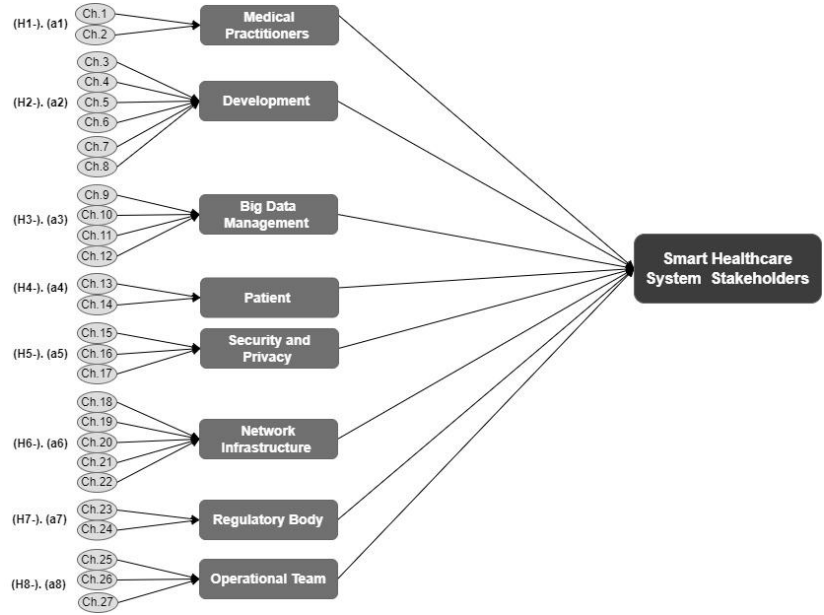

**Figure 2:** Proposed research model

### 3.2 Empirical research study design
This research study is conducted using the quantitative approach. The survey questionnaire was developed and analyzed in order to collect the responses from the respondents of each stakeholders working the domain of smart healthcare system. The total 85 responses were collected from the stakeholders. The questionnaire was distributed to stakeholders using the snowball technique and with the help of enumerator. The following subsequent sections describes the sampling method, survey questionnaire, and survey execution.

### 3.2.1 Sampling method
The survey-based studies usually comprised of two types of sampling method such as probabilistic sampling, and non-probabilistic sampling [88-90]. There are further six type of sampling method under the category of probabilistic sampling and non- probabilistic. These sampling methods are sorted in the descending order by considering the level of sample arbitrariness "(1) random sampling, (2) stratified sampling, (3) systematic sampling, (4) convenience sampling, (5) quota sampling, and (6) purposive sampling" [88-90]. The probabilistic sample is considered a systematic technique where every participant gets the equal amount of chance to be selected from the whole papulation. However, the probabilistic sampling is not always applicable, thus researchers take the advantage of non-probabilistic sampling where they itself take charge to select the sample. The probabilistic sampling and its subsequent types such as stratified sampling is usually adopted when there exists a large dataset thus cannot be used for this study as we have access to small dataset.

A-posteriori probability-based (systematic) sampling is deemed a method that imitate the concept of probability sampling. In A-posteriori probability-based (systematic) sampling, the data taken form companies are arranged into the group and then a sample is obtained from the collected data thus provides a chance to be selected for every participant. However, the participants would consider giving up reporting the honest opinion if their name or their organization's name are disclosed. Further, revealing the personal information could impose significant negative impact on the quality of project for which data is reported by participants.

Convenience sampling is a type of non-probabilistic sampling approach where "Subjects are selected because of their convenient accessibility to the researcher". All research studies consider testing the entire papulation ideal to get better result from the dataset, but this approach does not work in most of the cases. Therefore, researchers select the convenience sampling as this is considered the simplest techniques in the field of software engineering. We decided to use convenience sampling keeping in view the advantages and disadvantages of this technique.

### 3.2.2 Development of Survey Instrument

The initial survey draft was developed following the basic guidelines and experience reports by other researchers of the software engineering domain [88-90]. We then entered to the pilot testing phase and sent designed questionnaire to our industrial practitioner to make sure that the key terms used in the questionnaire survey was familiar to members of survey participants. The pilot testing is used to ensure the relevancy of terminologies used in the industry and research practitioners as study shows that terminologies used in the industry are slightly different from terminologies used by the research practitioners. Thus, the feedback from the industrial practitioners were analyzed and then helped us to finalize the survey questionaries.

The pilot testing helped us to finalize the questionnaire based on the feedback taken from the industry practitioners. The final draft of questionnaire survey comprised of 27 questions. The questionnaire used for the data collection is shown at https://doi.org/10.6084/m9.figshare.19636389. A section of questionnaire survey comprised of demographic information of the project, participant personal information, and companies' information. all other sections of the survey comprised of the questions related to know the significance of the of the identified key challenges CCF.

### 3.2.3 Recruitment of subjects and survey execution

The survey study was executed using the Google Form tools (docs.google.com/forms) and stored on the google drive giving access to all the participants. The survey study was approved by the research ethics departments of LUT University, in December 2021.

All the participants were contacted using the snowball techniques. We contacted the participant using all the available social media platform such as LinkedIn, Email, Facebook, ResearchGate, Twitter, etc. Once we get the response from the company participant, we further asked them to forward the questionnaire survey to their contacts as well. The ultimate goal was to maximize the number of members to take part in the research project. However, it is too obvious for us to determine that how many members were delivered the questionnaire survey and out of them how many responded (i.e., the response rate). We make sure not to send the multiple invitation (from different sources) to same respondent, we contacted using single point of platform for each participant or organization. Our plan included the target sectors, contact person, publicity schedule and status for each organization. All the survey participants were volunteer to provide the basic information and they can be withdrawn at any time.

There might be "duplicate entries" problem as multiple participants can work on the same completed or uncompleted project thus creating a problem where two or more data points represent the same unit. The research project is so diverse thus need to contact practitioners of each stakeholder of the working for implementation of smart healthcare ecosystem. However, the nature of our research project let us consider the duplicate entities in our data set. We believe that different project stakeholder could have different kind of opinion working under the same project.

There might be another problem where participant may not answer all the questions of questionnaire survey as it is possible in almost any online software engineering survey [91, 92] . However, the partial answer are often considered inadequate to conclude the result of the research study. For replicability and also to enable other researchers to conduct further analysis on our dataset, we have made our raw dataset available in an online resource (https://figshare.com/s/b2c785ffbd696dd6f4f7).

## 4 Results and Analysis

The results and analysis of questionnaire survey study are discussed in this section.

### 4.1 Profiles and demographics

The smart healthcare is a kind of ecosystem where serval stakeholders participate to make the whole ecosystem successful. These stakeholders can be medical practitioners, patients, operational team, regulatory bodies, development (manufacturer), network infrastructure provider, storage service provider, and insurance organizations. However, it is necessary to make sure the survey participants should have experiences of working with smart healthcare systems. Thus, we targeted the relevant population by verifying their profiles and demographics with respect our study objectives. For example, we contacted the manufacturer of the smart devices, analyzed the participant profile and their demographic to access the relevancy related to our survey study objective. The frequency of participants from each category of stakeholders are presented in Figure 3.

Since various stakeholders take part in the implementation of the smart healthcare ecosystem. The number of participants were analyzed based on the gender and we observed that 65% were male and 35% were female. We noted the maximum number of females participants are from medical practitioners' stakeholder's category.

The survey respondents were inquired about their working country with the aim to check the frequency of survey participants in terms of developed and underdeveloped countries. Thus, we observed that most of our survey respondents are from USA, China, UAE and India. All the participants were further inquired about their work experiences. The results show the majority of survey participants experiences is between 5 to 14 years. We also observed some of the respondents have experience more than 20 years. The mean and median values of survey participants shows that there is young and experienced pool of respondents.

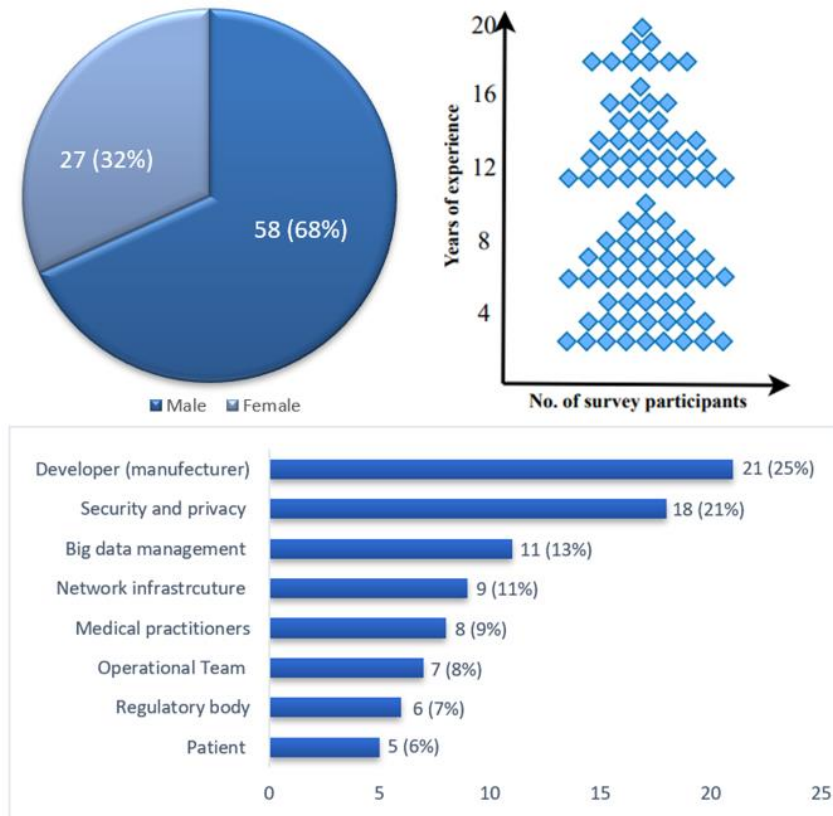

**Figure 3:** Demographics of survey participants

### 4.2 Data normality
We applied the skewness and kurtosis tests in order to obtain the normality of the collected data from the respondents. The result of skewness and kurtosis revealed that the data lies approximately in the normal distribution. The value lies between +1 to -1 as well as +2 to -2 is considered as the acceptable normal distribution values according to George and Mallery [93].

### 4.3 Internal consistency
The internal consistency or reliability can be measured using the Cronbach's Alpha and Composite reliability. All eight variables were tested using Cronbach's Alpha and Composite reliability and finally achieved satisfactory value presented in the Table 1, the value 70 or above is considered the satisfactory value achieved through Cronbach's Alpha and Composite reliability according Henseler et al. [94].

Table 1: Reliability Statistics

| Variables | Cronbach's Alpha | Composite Reliability | Original No. of Items | Final No. of Items |
|---|---|---|---|---|
| Medical Practitioners | 0.809 | 0.833 | 2 | 2 |
| Development | 0.973 | 0.981 | 5 | 5 |
| Big Data Management | 0.899 | 0.907 | 4 | 4 |
| Patient | 0.751 | 0.822 | 2 | 2 |
| Security and Privacy | 0.989 | 0.998 | 3 | 3 |
| Network Infrastructure | 0.954 | 0.961 | 5 | 5 |
| Regulatory Bodies | 0.855 | 0.891 | 2 | 2 |
| Operational Team | 0.901 | 0.922 | 3 | 3 |

### 4.4 Indicator reliability

Outer loading is used to measure the indicator reliability. The value equal to 0.7 or above is considered as satisfactory similar to internal consistency [94]. All the challenges achieved the satisfactory results from the indicator reliability test. The results are presented in the Table 2.

Table 2: Indicator Reliability through Outer Loadings.

|      | MP    | Dev   | BDM   | P     | SP    | NI    | RB    | OP    |
|------|-------|-------|-------|-------|-------|-------|-------|-------|
| MP1  | 0.786 |       |       |       |       |       |       |       |
| MP2  | 0.744 |       |       |       |       |       |       |       |
| Dev1 |       | 0.899 |       |       |       |       |       |       |
| Dev2 |       | 0.876 |       |       |       |       |       |       |
| Dev3 |       | 0.884 |       |       |       |       |       |       |
| Dev4 |       | 0.901 |       |       |       |       |       |       |
| Dev5 |       | 0.956 |       |       |       |       |       |       |
| Dev6 |       | 0.913 |       |       |       |       |       |       |
| BDM1 |       |       | 0.781 |       |       |       |       |       |
| BDM2 |       |       | 0.791 |       |       |       |       |       |
| BDM3 |       |       | 0.872 |       |       |       |       |       |
| BDM4 |       |       | 0.892 |       |       |       |       |       |
| P1   |       |       |       | 0.798 |       |       |       |       |
| P2   |       |       |       | 0.875 |       |       |       |       |
| SP1  |       |       |       |       | 0.985 |       |       |       |
| SP2  |       |       |       |       | 0.958 |       |       |       |
| SP3  |       |       |       |       | 0.957 |       |       |       |
| NI1  |       |       |       |       |       | 0.817 |       |       |
| NI2  |       |       |       |       |       | 0.872 |       |       |
| NI3  |       |       |       |       |       | 0.877 |       |       |
| NI4  |       |       |       |       |       | 0.981 |       |       |
| NI5  |       |       |       |       |       | 0.875 |       |       |
| RB1  |       |       |       |       |       |       | 0.834 |       |
| RB2  |       |       |       |       |       |       | 0.845 |       |
| OP1  |       |       |       |       |       |       |       | 0.856 |
| OP2  |       |       |       |       |       |       |       | 0.826 |
| OP3  |       |       |       |       |       |       |       | 0.708 |

**4.5 Multicollinearity**
In a multiple regression equation, the degree of correlation between two independent variables is known as multicollinearity. The VIF value is used to measure the multicollinearity. The value equal or higher than 10 is reveal the inconsistent in multicollinearity [95]. All the VIF values against their variables is consistent and shows no multicollinearity issue. The values are presented in the Table 3.

Table 3: Multicollinearity Values

| Independent Variables | Smart healthcare system | Process effectiveness |
|---|---|---|
| Medical Practitioners | 3.776 | 1.000 |
| Development | 5.324 | |
| Big Data Management | 4.898 | |
| Patient | 3.746 | |
| Security and Privacy | 6.541 | |
| Network Infrastructure | 5.231 | |
| Regulatory Bodies | 2.763 | |
| Operational Team | 5.131 | |

## 4.6 Validity analysis

Validity of data is used to test precisely what is intended to measure. Two methods mainly convergent and discriminant validity analysis is considered to measure the validity of actual data. The average variance extracted (AVE) method is used to measure the convergent validity. The AVE value above 0.5 is intended to be satisfactory [94]. All the variables were tested by convergent validity method. The results are acceptable revealed by convergent validity test.

Secondly, the discriminant validity is measured through PSL is Forner-Lorcker criterion. The discriminant validity is assumed to be acceptable if the square root of the AVE is higher than the correlation among the variable according to Forner-Lorcker [96]. Thus, the result of all variables obtained through the discriminant validity is considered as satisfactory and fulfill the criteria according to [97]. The result is presented in the Table 4.

According to the results given in Table 5, all the variables fulfill the satisfactory discrimination validity value. The diagonal value in green shape presents the square roots of the corresponding variables, and the value below the square root presents the correlation between variables. Considering Layman terms, the diagonal values should be greater than the values below the diagonal values.

Table 4: AVE Values of Latent Variables

| Independent Variables | Average Variance Extracted (AVE) |
|---|---|
| Medical Practitioners | 0.7532 |
| Development | 0.976 |
| Big Data Management | 0.846 |
| Patient | 0.712 |
| Security and Privacy | 0.985 |
| Network Infrastructure | 0.911 |
| Regulatory Bodies | 0.789 |
| Operational Team | 0.899 |

Table 5: Fornell-Larcker Criterion for Discriminant Validity

| | 1 | 2 | 3 | 4 | 5 | 6 | 7 | 8 |
|---|---|---|---|---|---|---|---|---|
| Medical Practitioners (1) | 0.857 | - | - | - | - | - | - | - |
| Development (2) | 0.950 | 0.897 | - | - | - | - | - | - |
| Big Data Management (3) | 0.856 | 0.772 | 0.856 | - | - | - | - | - |
| Patient (4) | 0.723 | 0.755 | 0.790 | 0.841 | - | - | - | - |
| Security and Privacy (5) | 0.978 | 0.985 | 0.990 | 0.831 | 0.768 | - | - | - |
| Network Infrastructure (6) | 0.841 | 0.852 | 0.855 | 0.855 | 0.851 | 0.798 | - | - |
| Regulatory Bodies (7) | 0.641 | 0.721 | 0.755 | 0.771 | 0.832 | 0.891 | 0.781 | - |
| Operational Team (8) | 0.811 | 0.758 | 0.768 | 0.821 | 0.613 | 0.712 | 0.751 | 0.611 |

## 4.7 Model fitness

There are several measures in PSL such as SRMR, NFI, and rms_Theta is used for model fitness. The bootstrap based Standard Root Mean Square (SRMR) is considered as a best alterative of chi-square [98, 99]. The value greater than 0.1 is considered as an inconsistent and below than 0.1 is considered as consistent in SRMR. All the obtained values are below than 0.1 satisfying the fitness of model. Further the value of NFI is also greater than 0.5 and less than 1 satisfying fit for the model. The value of rms_Theta is also closer to zero which is considered good fit.

Table 6: Model Fitness

|  | Saturated Model | Estimated Model |
| --- | --- | --- |
| SRMR | 0.068 | 0.086 |
| Rms_Theta | 0.198 |  |

## 4.8 Structural model evaluation

The proposed hypotheses are tested in the second step of PLS-based findings using their effect and significance. The degree (estimate values), significance (P Values and T-statistics), and $R^2$ measure of the structural model are all evaluated via bootstrapping. If the T-statistics value is more than 1.96 (with a significance level of 5%) or if it is more than 1.65 (with a significance level of 10%), it is deemed significant. Each connection will be examined in the following subsections using parameter estimates (beta values) and T- statistics. The findings of the structural model evaluation are shown in Table 7. The findings of the causal structural model are shown in Fig. 4.

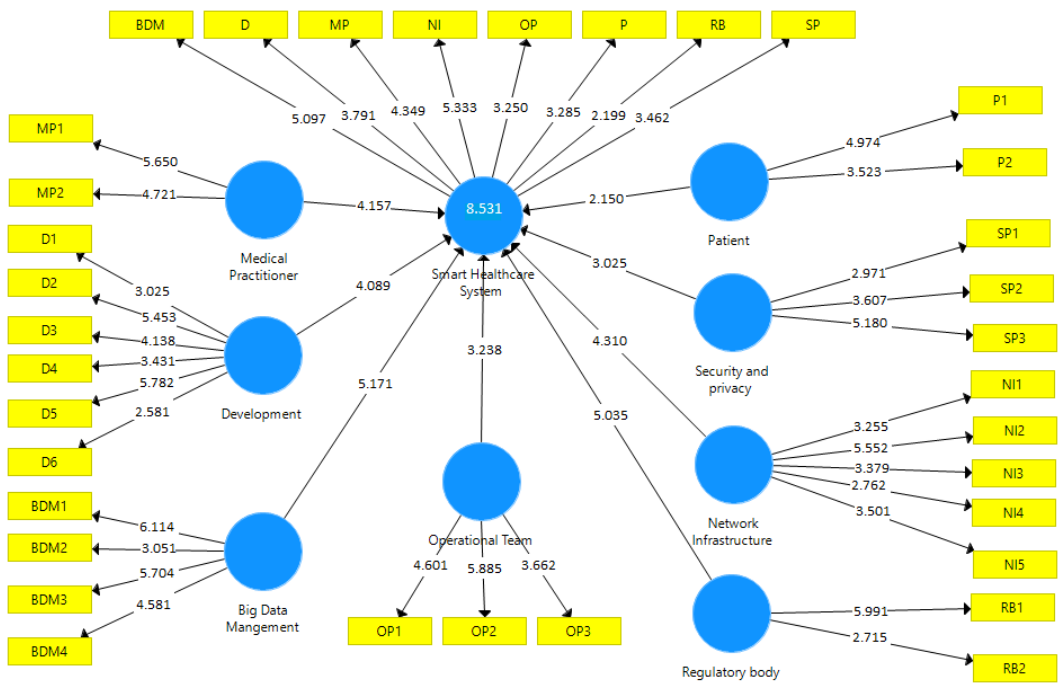

**Figure 4:** Structural Model Evaluation

## 4.9 Medical practitioners and smart healthcare system

The influence of medical practitioner's challenges on the smart healthcare system is significant as the P-value is: 0.000; T-value:4.259 and relatively beta value is 0.638. The results presented in the Table 7 shows that the each increasing number of challenges increase the Beta value that highlights the importance of hypothesis 1(H1).

### 4.10 Development (manufacturer) and smart healthcare system
The manufacturer is the key stakeholders in the development of smart healthcare system. Thus, the influence of manufacturer on the smart healthcare system is quite significant presented by T-value: 26.058 and the P-value: 0.000. Further, the beta value: 0.988 indicate the robust influence of manufacturer on the smart healthcare system. The value is increased by 0.988 with increase in each unit. Thus, manufacturer is negatively and significantly influence on the smart healthcare system.

### 4.11 Big data management and smart Healthcare system
Big data management has significant negative impact on the smart healthcare system if the challenges are not properly addressed. The result presented in the table 7 (T-value: 4.088; P-value: 0.000, beta value: 0.771) shows the significant impact on the healthcare system. The increase in the challenges in big data management stakeholder category decrease the beta value. Hence result support our hypothesis.

### 4.12 Patient and smart healthcare system
The intention of patients toward the use of smart healthcare system is significant as the T-value is 9.538 and P-value is 0.000 Furthermore, the beta values: 0.487 that indicate the patient's intention toward the improvement of smart healthcare system is significant thus supporting the hypothesis 2 (H2).

### 4.13 Security, Privacy and Smart healthcare system
The increasing number of smart devices for healthcare purpose are significantly becoming potent to several security and privacy vulnerabilities. The result presented in the table 7 (T-value: 9.932; P-value: 0.000, beta value: 0.991) shows the significant impact on the healthcare system. The increase in the challenges in big data management stakeholder category decrease the beta value. Hence result support our hypothesis.

### 4.14 Network Infrastructure and smart healthcare system
The results predicted that network infrastructure has negative influence on the smart healthcare system with (Beta value: 0.843; T-Value: 8.589: P-value:0.000). Therefore, the increase in the challenges in network infrastructure stakeholder will eventually decrease the beta value. Hence, support the proposed hypothesis 6 (H6).

### 4.15 Regulatory Body and smart healthcare system
Regulatory body poses significant influence on the successful implementation of smart healthcare system. However, the results in the table 9 (Beta value: 0.651; T value: 7.512: P-value: 0.000) predicted that with each increase in the challenge of regulatory body there is decrease in the beta value. Thus, the result support hypothesis 5 (H5).

### 4.16 Operational team and smart healthcare system
The result presented in the table 7 indicate that operational team has significant negative impact on the smart healthcare system with the (Beta value: 0.751: T-Value: 9.581: p-Value: 0.000). The beta value indicates that with the increase in each challenge of operational stakeholder the influence of smart healthcare decrease. By 0.751 unit. Therefore, results shows that operational team challenges negatively influence on the smart healthcare system supporting the hypothesis 4 (H4).

Table 7: Bootstrap Results for Causal Structural Model

| Parameters | Parameter Estimate | Sample Mean | Standard Deviation | T Statistics | P Values |
|---|---|---|---|---|---|
| Medical Practitioners → Smart healthcare system | 0.638 | 0.738 | 0.032 | 4.259 | 0.000 |
| Development → Smart healthcare system | 0.988 | 0.989 | 0.099 | 26.058 | 0.000 |
| Big data management → Smart healthcare system | 0.771 | 0.668 | 0.066 | 4.088 | 0.000 |
| Patient → Smart healthcare system | 0.487 | 0.487 | 0.051 | 9.538 | 0.000 |
| Security and Privacy → Smart healthcare system | 0.991 | 0.562 | 0.092 | 9.932 | 0.000 |
| Network Infrastructure → Smart healthcare system | 0.843 | 0.841 | 0.521 | 8.589 | 0.000 |
| Regulatory Bodies → Smart healthcare system | 0.651 | 0.521 | 0.052 | 7.512 | 0.000 |
| Operational Team → Smart healthcare system | 0.751 | 0.721 | 0.063 | 9.581 | 0.000 |

### 4.17 The $R^2$ for the endogenous latent variables

The coefficient of determinations ($R^2$) is used to check the variation that is introduced by independent variables with relation to the dependent variable. All the values of determined coefficient of determination ($R^2$) are presented in the Table 8. The analyzed results of $R^2$ between exogenous variables which include 8 key stakeholders, and the endogenous variable (Smart healthcare system) are substantial and significant as $R^2$=0.74 and T=32.623. This represents the need to address the key stakeholders' challenges for successful implication of the smart healthcare. Senapathi and Srinivasan[100] mentioned that if the R2 ≥ 0.75 is considered substantial.

Table 8: $R^2$ value for endogenous latent variables

| | Original Sa | N | SD | T Stat | P Val |
|---|---|---|---|---|---|
| Smart healthcare system | 0.794 | 0.800 | 0.024 | 32.623 | 0.000 |

## 5. Discussion

### 5.1 Medical practitioners and smart healthcare system

**Hypothesis (H1)-** (Medical practitioners need consideration for the successful implementation of smart healthcare).

To summarize, it is hypothesized that Ch.1 and Ch.2 negatively impact the medical practitioners. There is need to address both challenges for the successful implementation of smart healthcare.

The results of PSL showed that medical practitioners are a significant stakeholder category that could negatively impact on the smart healthcare system if challenges (Ch.1 and Ch.2) are not addressed. The result T-value: 4.259: P-value: 0.000 and beta value 0.638 revealed that by addressing the highlighted challenges of medical practitioner category, the beta value will decrease. The decrease in beta value shows the successfulness of medical practitioners of stakeholder's category.

The main cause behind the results can be noticed in the key challenges associated with medical practitioners. It has been observed that lake of authentication of medical practitioners possess significant negative impact on the smart healthcare system. There is no such system, which record the profile of medical practitioners i.e., the authentic work experience in the medical field, the core expertise, the number of patients medical practitioners have so far examined, the success and failure rate of those cases, expertise in using the modern equipment, and the novelty in his core domain. In addition to this the lake of certification for the use of state-of-the art medical devices is another significant challenge. Medical practitioners without having deep knowledge on the use of smart healthcare system may pose significant negative impact on the whole ecosystem. Therefore, lake of authentication and lake of certification make this stakeholder significantly important to address for the successful implementation of smart healthcare system.

### 5.2 Development (manufacturer) and smart healthcare system

**Hypothesis (H2):** manufacturer may pose negative influence on the smart healthcare system

The relationship between the manufacturer and healthcare system is significant. The identified challenges pose the negative influence on the successful implementation of smart healthcare system. The results of PSL revealed that T-value: 26.058 and the P-value: 0.000. The beta value 0.988 indicate the robust negative influence of manufacturer on the smart healthcare system. The value is decreased by 0.988 with increase in each challenge associated with the manufacturer category of smart healthcare system.

The manufacturer of smart healthcare devices may face several challenges such as lack of interoperability among the heterogeneous platforms and standards. The rise in the use of smart healthcare devices with having heterogeneous nature of standards could enforce the significant challenge related to security and privacy, Similarly, Lake of robust framework for scaling IoT infrastructure, Lake of software updates, Size limitation of smart devices, Lake of ethics considerations, Lake of self-learning and self-improvement methods for devices are the key challenges that manufacture stakeholder category need to address for successful implementation of smart healthcare system.

The literature and the empirical result revealed the significance of the manufacturer stakeholder. We agree that the current study's findings are valid and skewed toward the majority of research findings concerning the relation between manufacturers and smart healthcare systems.

### 5.3 Big Data Management and smart healthcare system

**Hypothesis (H3):** Big data issues can pose negative impact on the smart healthcare system

The result of the analysis revealed that big data management issue may negatively influence on the implementation of the smart healthcare system. The result such as T-value: 4.088; P-value: 0.000, beta value: 0.771showed that the beta value is decreased with the increase in the challenges associated with big data management stakeholder category.

The outcomes of this study support the authors' assertion that there is a positive correlation between big data management and smart healthcare systems.

Several authors have argued on different challenges that could occur due the huge amount of data being generated by the smart healthcare devices. some authors argue that the heterogenous nature of the data format such as XML and other format is the main cause of growing data and its management. The lake of framework for managing heterogenous data could negatively influence on the entire system. similarly, lake of techniques for extraction of clean, formatted, thorough, and precise data could be challenging in emergency cases.

### 5.4 Patients and smart healthcare system

**Hypothesis (H4):** Patients intention towards the use of healthcare system could negatively impact.

The research finding revealed that patient's intention toward the use of smart healthcare system could negatively influence. The T-value and P-value is 9.538 and 0.000 respectively. Furthermore, the beta value: 0.487 showed that with increase in the challenges in patients stakeholder category decrease the beta value proving the hypothesis H4 (H4).

The result of PLS supporting the finding of study reported in the literature. Patients have lake of trust on the use of smart healthcare devices as they are more concerned about their privacy of personal medical information to be disclosed, location, transparency, and accountability [11,12]. Similarly, illiterate patient is another challenge for implementation of smart healthcare ecosystem in the area where illiteracy rate is higher.

Therefore, the identified challenges are supporting hypothesis and results showed that these challenges negatively influencing and hindering the adoption of healthcare system.

### 5.5 Security, Privacy, and smart healthcare system

**Hypothesis (H5):** smart healthcare devices are more prone to security and privacy threats that could negatively impact on the healthcare system

The research finding showed that security and privacy is the main stakeholder category that significantly posing negative influence on the smart healthcare system. The T-value and P-value is 9.932 and 0.000 respectively. Furthermore, the beta value: 0.991 showed that with increase in the challenges in patients stakeholder category decrease the beta value proving the hypothesis H5 (H5).

The literature reported several security and privacy challenges that are negatively influencing the on the development of healthcare system. The heterogenous nature of architecture offers several layer-based security and privacy threats such as node capture and cloning, Eavesdropping, Jamming attack, resource depletion attack, relay attack. Similarly, these smart devices are lacking in the state-of-the art vulnerability assessments techniques due to small power and size. Further, the heterogeneous nature of these devices impeding researcher to develop the automated detection and recovery of vulnerabilities [17]. The smart devices further lacking in the framework that could detect the malicious devices in the network.

Therefore, we believe that security and privacy is the significant stakeholder category that is negatively influencing the smart healthcare system.

### 5.6 Network infrastructure and smart healthcare system

**Hypothesis (H6):** lack of management of network infrastructure could negatively influence on the smart healthcare system.

The results predicted that network infrastructure has negative influence on the smart healthcare system with (Beta value: 0.843; T-Value: 8.589: P-value: 0.000). Therefore, the increase in the challenges in network infrastructure stakeholder will eventually decrease the beta value. Hence, support the proposed hypothesis 6 (H6).

Network infrastructure is the key stakeholder category of smart healthcare system. However, several challenges could pose significant impact on the healthcare system. huge amount of data is being collected over the could thus lake of data processing cloud service provider could negatively impact on the healthcare system. similarly, lake of reliable connectivity in the developed countries could be challenge in implementation of healthcare system. The lack of security standards in cloud could cause several attacks and unavailability of medical history of the patients. Similarly, client management is another challenge.

### 5.7 Regulatory body and smart healthcare system

**Hypothesis (H7):** Lake of globally recognized regulations could negatively influence on the smart healthcare system

Regulatory body poses significant influence on the successful implementation of smart healthcare system. However, the results in the table 9 (Beta value: 0.651; T value: 7.512: P-value: 0.000) predicted that with each increase in the challenge of regulatory body there is decrease in the beta value. Thus, the result support hypothesis 8 (H8).

The regulatory bodies such as FDA and HIPPA has developed their own set of rules for the approval of smart healthcare devices. These bodies keep changing their rules that eventually impact on the development of these devices. Furthermore, nearly all developed states have formulated their own individual regulatory standards on the manufacturing and use of smart healthcare devices. However, manufacture of the smart devices needs to compliance with the rules of these regulatory bodies. Thus, change in regulation and lake of global consensus over the development and use of smart healthcare is challenging for the success of modern healthcare system.

### 5.8 operational team and smart healthcare system

*Hypothesis-8 (H8): Implementation of smart devices and network in the hospital require experienced operation team. Thus, a need of planning required to address the [Ch.25 -Ch27] challenges that could impose negative impact over the ecosystem.*

The result presented in the table 9 indicate that operational team has significant negative impact on the smart healthcare system with the (Beta value: 0.751: T-Value: 9.581: p-Value: 0.000). The beta value indicates that with the increase in each challenge of operational stakeholder the influence of smart healthcare decreases relatively by 0.751 unit. Therefore, results shows that operational team challenges negatively influence on the smart healthcare system supporting the hypothesis 8 (H8).

Smart healthcare ecosystem is rapidly growing mainly in the developed countries. The increase in the use of concept demands highly professional operational team that can implement, troubleshoot, and evaluate network or device. The Lake of smart healthcare infrastructure management professionals and workflow disruption during implementation could be serious for the whole system success. Furthermore, lake of security professionals that manage the network cybersecure is another challenge.

### 6. Study Implications and Future Directions
This research study has significant implication for researchers and industrial practitioners in order to make smart healthcare system adoptable and successful.

**Researchers:** this study provides the overview of state-of-the-art literature conducted in the context of smart healthcare system stakeholders. This study elaborates the importance of smart healthcare systems stakeholders and identify the critical challenging factors against each stakeholder. The result of this study serves a body of knowledge to research community to develop the new and effective roadmaps for the success and progression of smart healthcare systems. Furthermore, empirical study also conducted with experts with the aim to check the criticality of the identified stakeholders and their related challenges. The empirical results encourage the researchers as this study presents the real-world problems facing by smart healthcare systems stakeholders. To summarize, our study results give a positive push to research community to research the smart health systems care areas in the context to develop the strategies to cover the systems from all sides of stakeholders.

**Practitioners**: The deep overview of the state-of-the-art literature and practices will help the practitioners to understand the importance of each smart healthcare stakeholders. Furthermore, the practitioners can understand the critical challenges facing by each stakeholder to give their best to make the smart healthcare system successful and effective for practices. Considering the study results and analysis, the industry practitioners could revise their policies for stakeholders with the intent to make the implication of smart healthcare system successful. In addition, the identified challenging factors will assist the industry experts to develop the new strategies and effective policies in order to address the identified challenges which could help the better input from each stakeholder.

**Future direction**

In future, we will extend this study by developing a prioritization-based taxonomy of the smart healthcare stakeholders' challenges, by using fuzzy analytical hierarchy approach. The develop prioritization-based taxonomy will help the industry experts to address the multicriteria decision making problems. By conducting the case studies with real-world practitioners, the practical robustness of the proposed prioritization-based taxonomy will be assessed.

Moreover, we also plan to conduct interview study with practitioners to identify the success factors by applying grounded theory approach. Then, considering the success and challenging factors, we will develop a readiness model that could provide the roadmap for industry experts to develop the effective policies for stakeholders' involvement in each phase (development of implementation) of smart healthcare system execution. The practical implacability of the readiness model will also be measured by conducting the case studies with different industries involved in smart healthcare process development and execution.

## 7. Threats to validity

This section describes the potential threat to validity of the conducted research study and how they are addressed by following the guideless proposed by [101-103]. The potential threats are categorized into four types as: internal validity, external validity, construct validity, and conclusion validity[104, 105].

**Internal Validity** refers to the variables that ultimately effect on the result and analysis derived by different means. The study is based on the formal literature review and an empirical investigation. We have explored the key stakeholders and their respective challenges from the literature and empirically validated. Therefore, one possible threat could be missing the primary studies related to the scope of this study research. First, this omission is not systematic [106, 107]. Secondly, we tried to mitigate this threat by empirically validating literature findings with industry experts.

The respondents usually keep the questionnaire survey and respond very late that effect on sample size. This threat was mitigated by giving them time slots of two weeks.

Furthermore, a total 85 participant from all stakeholders' categories had great interest in their respective roles that could raise the issue of biasness. However, the experience of each participant was different from low to high as contribute to mitigate this prejudice.

**External Validity** refers to the generalization of the results and findings of the conducted research. The empirically collected data is particularly based on the eight key stakeholders and their respective challenges. A possible threat to this study could be whether the shared data collection instrument was same with 85 participants. This threat was mitigated by inviting the participants of their respective category and a single concrete questionnaire was shared among them. The questionnaire was comprised of all stakeholders and their respective challenges. The participants were requested to fill their related sections of the questionnaire.

Another threat to this study is, whether a total 85 participant are enough to generalize the results as small size of sampling could be a limitation. Based on the other empirical studies conducted in other engineering domain[108-110], the collected sample size is good enough to generalize the study results. We then observed the participant positions they were working in the project as different practitioners in the same project have different experiences and opinion [111]. Thus, there are kind of possibility where are single participants has more than one position and therefore respondent were permitted to choose more than one position.

**Construct Validity** refers to the comprehensiveness of the developed survey instrument. To get the pure insight, it is necessary the definition of research objective and the statements of the variables should be understandable. This threated alleviated by conducting the pilot study with experts. Furthermore, to mitigate this threat by adding a space to write participant own answer by choosing "other" option.

**Reliability** refers to the validity of our research findings. This threat has been addressed by discussing data collection, analyses and synthesis process with research team, and research advisor. We also follow the research ethics rules according to the LUT University guidelines.

## 8. Conclusions

Smart healthcare system becomes an important part of this modern era. As for individual users, smart healthcare can facilitate better health self-management, timely and appropriate medical services that can be accessed when needed, and the content of medical services will be more personalized. The smart healthcare system assists practitioners in making quicker decisions that have a higher priority to be hospitalized, which makes for the effective allocation of scarce medical resources. However, the smart healthcare system is vulnerable to several challenges that are negatively influencing the successful adoption of this ecosystem. The key objective of this research work is to elaborate the importance of key stakeholders and their respective challenges from the state-of-the art literature. We have identified the eight stakeholders i.e., "medical practitioner, patients, operational team, regulatory bodies, Developer (manufacturer), network infrastructure provider, storage service provider, and cybersecurity", reported in literature. We further identified the 27 challenges that could negatively influence the workability and participate-ability of stakeholders in the development and implications of smart healthcare systems. Based on the literature discussion, we proposed hypothesis as the identified challenging factors have negative impact on the stakeholder's consideration in smart healthcare system, and the ineffective stakeholder participation, have negative impact on the successful development and execution of smart healthcare system. In second step, the identified stakeholders and challenges were further analyzed with industrial practitioners and researchers. A questionnaire survey was distributed, and a total of 85 complete responses were collected. The collected responses were further analyzed using smart PLS approach. The result of analysis shows that the stakeholder category and their respective challenges negatively influencing from development to adoption of smart healthcare system. According to the results, the developer (manufacturer) category is the most significant among all stakeholders' categories. We believe that the finding of this research study would assist research community to focus on identified challenges in order to make sure the better participation of all smart healthcare systems stakeholders, that could significantly contribute to make this system successful.

## References

1. Welfare, A.I.o.H.a. *Australia's Health*. 2014.
2. Miller, I.F., et al., *Disease and healthcare burden of COVID-19 in the United States.* Nature Medicine, 2020. **26**(8): p. 1212-1217.
3. Esteva, A., et al., *Dermatologist-level classification of skin cancer with deep neural networks.* nature, 2017. **542**(7639): p. 115-118.
4. (WHO), W.H.O., *Global Health and Aging*. 2021.
5. HealthAffairs, *National health expenditure projections, 2018–27: economic and demographic trends drive spending and enrollment growth*. 2019.
6. Pramanik, M.I., et al., *Smart health: Big data enabled health paradigm within smart cities.* Expert Systems with Applications, 2017. **87**: p. 370-383.
7. Chang, S.-H., et al., *A context-aware, interactive M-health system for diabetics.* IT professional, 2016. **18**(3): p. 14-22.

8. Harimoorthy, K. and M. Thangavelu, *Cloud-assisted Parkinson disease identification system for remote patient monitoring and diagnosis in the smart healthcare applications.* Concurrency and Computation: Practice and Experience, 2021. **33**(21): p. e6419.
9. Kumar, N. *IoT architecture and system design for healthcare systems.* in *2017 International Conference on Smart Technologies for Smart Nation (SmartTechCon).* 2017. IEEE.
10. Malhotra, P., et al., *Internet of things: Evolution, concerns and security challenges.* Sensors, 2021. **21**(5): p. 1809.
11. Sadique, K.M., R. Rahmani, and P. Johannesson, *Towards security on internet of things: applications and challenges in technology.* Procedia Computer Science, 2018. **141**: p. 199-206.
12. Atlam, H.F. and G.B. Wills, *IoT security, privacy, safety and ethics*, in *Digital Twin Technologies and Smart Cities.* 2020, Springer. p. 123-149.
13. Yuan, S., R.H. Rao, and S. Upadhyaya, *Emerging issues for education in E-discovery for electronic health records.* Security Informatics, 2015. **4**(1): p. 1-7.
14. Kute, S.S., A.K. Tyagi, and S. Aswathy, *Security, Privacy and Trust Issues in Internet of Things and Machine Learning Based e-Healthcare*, in *Intelligent Interactive Multimedia Systems for e-Healthcare Applications.* 2022, Springer. p. 291-317.
15. Atiyah, R.F. and I. Al-Mejibli, *Security and Privacy in IoT Healthcare System: A systematic review.* Journal of Al-Qadisiyah for computer science and mathematics, 2022. **14**(1): p. Page 15–23-Page 15–23.
16. Panda, A. and S. Mohapatra, *Online healthcare practices and associated stakeholders: review of literature for future research agenda.* Vikalpa, 2021. **46**(2): p. 71-85.
17. Marshal, R., K. Gobinath, and V.V. Rao. *Proactive Measures to Mitigate Cyber Security Challenges in IoT based Smart Healthcare Networks.* in *2021 IEEE International IOT, Electronics and Mechatronics Conference (IEMTRONICS).* 2021. IEEE.
18. Cheen, M.H.H., et al., *Prevalence of and factors associated with primary medication non-adherence in chronic disease: a systematic review and meta-analysis.* International journal of clinical practice, 2019. **73**(6): p. e13350.
19. Aggarwal, R., et al., *Rural-urban disparities: diabetes, hypertension, heart disease, and stroke mortality among black and white adults, 1999-2018.* Journal of the American College of Cardiology, 2021. **77**(11): p. 1480-1481.
20. Mansour, R.F., et al., *Artificial intelligence and Internet of Things enabled disease diagnosis model for smart healthcare systems.* IEEE Access, 2021. **9**: p. 45137-45146.
21. Agiwal, M., N. Saxena, and A. Roy, *Towards connected living: 5G enabled internet of things (IoT).* IETE Technical Review, 2019. **36**(2): p. 190-202.
22. Bush, A.A., A. Tiwana, and H. Tsuji, *An empirical investigation of the drivers of software outsourcing decisions in Japanese organizations.* Information and Software Technology, 2008. **50**(6): p. 499-510.
23. Algarni, A., *A survey and classification of security and privacy research in smart healthcare systems.* IEEE Access, 2019. **7**: p. 101879-101894.
24. Benzarti, S., B. Triki, and O. Korbaa. *A survey on attacks in Internet of Things based networks.* in *2017 International conference on engineering & MIS (ICEMIS).* 2017. IEEE.
25. Sadeeq, M.M., et al., *IoT and Cloud computing issues, challenges and opportunities: A review.* Qubahan Academic Journal, 2021. **1**(2): p. 1-7.
26. Ahmad, W., et al., *Cyber Security in IoT-Based Cloud Computing: A Comprehensive Survey.* Electronics, 2022. **11**(1): p. 16.
27. Saadia, D., *Integration of cloud computing, big data, artificial intelligence, and internet of things: Review and open research issues.* International Journal of Web-Based Learning and Teaching Technologies (IJWLTT), 2021. **16**(1): p. 10-17.
28. Zeadally, S., et al., *Smart healthcare: Challenges and potential solutions using internet of things (IoT) and big data analytics.* PSU research review, 2019.

29. Mubashar, M., et al., *Instadoc: doctors' appointment and management system*. 2017, University of Management and Technology Lahore.
30. catCert, *Identity and capability management in eHealth: the CATCert approach*. 2017.
31. De Grood, C., et al., *Adoption of e-health technology by physicians: a scoping review*. Journal of multidisciplinary healthcare, 2016. **9**: p. 335.
32. Luyten, J. and W. Marneffe, *Examining the acceptance of an integrated Electronic Health Records system: Insights from a repeated cross-sectional design*. International Journal of Medical Informatics, 2021. **150**: p. 104450.
33. Pan, J., et al., *Exploring behavioural intentions toward smart healthcare services among medical practitioners: A technology transfer perspective*. International Journal of Production Research, 2019. **57**(18): p. 5801-5820.
34. Qiu, T., et al., *How can heterogeneous internet of things build our future: A survey*. IEEE Communications Surveys & Tutorials, 2018. **20**(3): p. 2011-2027.
35. Rana, B., Y. Singh, and P.K. Singh, *A systematic survey on internet of things: Energy efficiency and interoperability perspective*. Transactions on Emerging Telecommunications Technologies, 2021. **32**(8): p. e4166.
36. Newman, J., *Why the Internet of things might never speak A common language*. App Economy, Fast Company, 2016.
37. Kollolu, R., *A Review on Wide Variety and Heterogeneity of IoT Platforms*. The International journal of analytical and experimental modal analysis, analysis, 2020. **12**: p. 3753-3760.
38. Dimitrov, D.V., *Medical internet of things and big data in healthcare*. Healthcare informatics research, 2016. **22**(3): p. 156-163.
39. Lee, E., et al., *A Survey on Standards for Interoperability and Security in the Internet of Things*. IEEE Communications Surveys & Tutorials, 2021. **23**(2): p. 1020-1047.
40. Shah, R. and A. Chircu, *IOT and ai in healthcare: A systematic literature review*. Issues in Information Systems, 2018. **19**(3).
41. Bahl, M., et al., *High-risk breast lesions: a machine learning model to predict pathologic upgrade and reduce unnecessary surgical excision*. Radiology, 2018. **286**(3): p. 810-818.
42. Bakar, N.A., W.M.W. Ramli, and N.H. Hassan, *The internet of things in healthcare: an overview, challenges and model plan for security risks management process*. Indonesian Journal of Electrical Engineering and Computer Science (IJEECS), 2019. **15**(1): p. 414-420.
43. AboBakr, A. and M.A. Azer. *IoT ethics challenges and legal issues*. in *2017 12th International Conference on Computer Engineering and Systems (ICCES)*. 2017. IEEE.
44. Sowmiya, B., et al., *A survey on security and privacy issues in contact tracing application of COVID-19*. SN computer science, 2021. **2**(3): p. 1-11.
45. Shahriari, K. and M. Shahriari. *IEEE standard review—Ethically aligned design: A vision for prioritizing human wellbeing with artificial intelligence and autonomous systems*. in *2017 IEEE Canada International Humanitarian Technology Conference (IHTC)*. 2017. IEEE.
46. Jammula, M., V.M. Vakamulla, and S.K. Kondoju, *Performance Evaluation of Lightweight Cryptographic Algorithms for Heterogeneous IoT Environment*. Journal of Interconnection Networks, 2022: p. 2141031.
47. Al-Husainy, M.A.F., B. Al-Shargabi, and S. Aljawarneh, *Lightweight cryptography system for IoT devices using DNA*. Computers & Electrical Engineering, 2021. **95**: p. 107418.
48. Wang, J., et al. *An energy harvesting chain model for wireless-powered IoT networks*. in *2018 10th International Conference on Wireless Communications and Signal Processing (WCSP)*. 2018. IEEE.
49. Schöffel, M., et al., *Secure IoT in the Era of Quantum Computers—Where Are the Bottlenecks?* Sensors, 2022. **22**(7): p. 2484.
50. Padilla, F.J.A., et al. *The future of IoT software must be updated*. in *IAB Workshop on Internet of Things Software Update (IoTSU)*. 2016.

51. Gupta, H. and P.C. Van Oorschot. *Onboarding and software update architecture for IoT devices*. in *2019 17th International Conference on Privacy, Security and Trust (PST)*. 2019. IEEE.
52. Pandey, A.K., B. Rajendran, and V. Roshni, *AutoAdd: Automated Bootstrapping of an IoT Device on a Network.* SN Computer Science, 2020. **1**(1): p. 1-5.
53. Helu, M., et al., *Scalable data pipeline architecture to support the industrial internet of things.* CIRP Annals, 2020. **69**(1): p. 385-388.
54. Balalaie, A., A. Heydarnoori, and P. Jamshidi, *Microservices architecture enables devops.* London: Sharif University of Technology, 2014.
55. Fawzy, D., S.M. Moussa, and N.L. Badr, *The Internet of Things and Architectures of Big Data Analytics: Challenges of Intersection at Different Domains.* IEEE Access, 2022.
56. Mothe, R., et al., *A Review on Big Data Analytics in Internet of Things (IoT) and Its Roles, Applications and Challenges*, in *ICDSMLA 2020*. 2022, Springer. p. 765-773.
57. Ahmed, E., et al., *The role of big data analytics in Internet of Things.* Computer Networks, 2017. **129**: p. 459-471.
58. Shahid, A., T.-A.N. Nguyen, and M.-T. Kechadi, *Big data warehouse for healthcare-sensitive data applications.* Sensors, 2021. **21**(7): p. 2353.
59. Anagnostopoulos, I., S. Zeadally, and E. Exposito, *Handling big data: research challenges and future directions.* The Journal of Supercomputing, 2016. **72**(4): p. 1494-1516.
60. Sharma, N. and R. Rohilla. *Blockchain Based Electronic Health Record Management System for Data Integrity*. in *Proceedings of International Conference on Computational Intelligence*. 2022. Springer.
61. Vaiyapuri, T., A. Binbusayyis, and V. Varadarajan, *Security, privacy and trust in iomt enabled smart healthcare system: A systematic review of current and future trends.* International Journal of Advanced Computer Science and Applications, 2021. **12**(2): p. 731-737.
62. Esha, N.H., et al. *Trust IoHT: a trust management model for internet of healthcare things*. in *Proceedings of International Conference on Data Science and Applications*. 2021. Springer.
63. Rohokale, V.M., N.R. Prasad, and R. Prasad. *A cooperative Internet of Things (IoT) for rural healthcare monitoring and control*. in *2011 2nd international conference on wireless communication, vehicular technology, information theory and aerospace & electronic systems technology (Wireless VITAE)*. 2011. IEEE.
64. Karamali, M., M. Yaghoubi, and A. Parandeh, *Scientific Mapping of Papers Related to Health Literacy Using Co-Word Analysis in Medline.* Iranian Journal of Health Education and Health Promotion, 2021: p. 41-50.
65. Yao, X., et al., *Security and privacy issues of physical objects in the IoT: Challenges and opportunities.* Digital Communications and Networks, 2021. **7**(3): p. 373-384.
66. Lombardi, M., F. Pascale, and D. Santaniello, *Internet of things: A general overview between architectures, protocols and applications.* Information, 2021. **12**(2): p. 87.
67. Mohanty, J., et al., *IoT security, challenges, and solutions: a review.* Progress in Advanced Computing and Intelligent Engineering, 2021: p. 493-504.
68. Samizadeh Nikoui, T., et al., *Internet of Things architecture challenges: A systematic review.* International Journal of Communication Systems, 2021. **34**(4): p. e4678.
69. Newaz, A.I., et al. *Healthguard: A machine learning-based security framework for smart healthcare systems*. in *2019 Sixth International Conference on Social Networks Analysis, Management and Security (SNAMS)*. 2019. IEEE.
70. Anand, P., et al., *IoT vulnerability assessment for sustainable computing: threats, current solutions, and open challenges.* IEEE Access, 2020. **8**: p. 168825-168853.
71. Yu, T., et al. *Handling a trillion (unfixable) flaws on a billion devices: Rethinking network security for the internet-of-things*. in *Proceedings of the 14th ACM Workshop on Hot Topics in Networks*. 2015.
72. Rayes, A. and S. Salam, *Internet of things security and privacy*, in *Internet of Things from Hype to Reality*. 2022, Springer. p. 213-246.

73. Hussain, F., et al., *A framework for malicious traffic detection in IoT healthcare environment.* Sensors, 2021. **21**(9): p. 3025.
74. Hafeez, I., et al., *IoT-KEEPER: Detecting malicious IoT network activity using online traffic analysis at the edge.* IEEE Transactions on Network and Service Management, 2020. **17**(1): p. 45-59.
75. Haji, L.M., et al., *Dynamic resource allocation for distributed systems and cloud computing.* TEST Engineering & Management, 2020. **83**(May/June 2020): p. 22417-22426.
76. Sonune, S., et al. *Issues in IoT healthcare platforms: A critical study and review.* in *2017 International Conference on Intelligent Computing and Control (I2C2).* 2017. IEEE.
77. Zhang, Y., et al., *Ensuring attribute privacy protection and fast decryption for outsourced data security in mobile cloud computing.* Information Sciences, 2017. **379**: p. 42-61.
78. Deebak, B.D. and A.-T. Fadi, *Lightweight authentication for IoT/Cloud-based forensics in intelligent data computing.* Future generation computer systems, 2021. **116**: p. 406-425.
79. Zagan, E. and M. Danubianu. *Cloud DATA LAKE: The new trend of data storage.* in *2021 3rd International Congress on Human-Computer Interaction, Optimization and Robotic Applications (HORA).* 2021. IEEE.
80. Hussein, A.F., et al., *An automated remote cloud-based heart rate variability monitoring system.* IEEE access, 2018. **6**: p. 77055-77064.
81. Tissaoui, A. and M. Saidi. *Uncertainty in IoT for Smart Healthcare: Challenges, and Opportunities.* in *International Conference on Smart Homes and Health Telematics.* 2020. Springer.
82. Leavitt, N., *Is cloud computing really ready for prime time.* Growth, 2009. **27**(5): p. 15-20.
83. Rimal, B.P., et al., *Architectural requirements for cloud computing systems: an enterprise cloud approach.* Journal of Grid Computing, 2011. **9**(1): p. 3-26.
84. Cote, R.A. *Architecture of SNOMED: its contribution to medical language processing.* in *Proceedings of the Annual Symposium on Computer Application in Medical Care.* 1986. American Medical Informatics Association.
85. Brönneke, J.B., et al., *Regulatory, Legal, and Market Aspects of Smart Wearables for Cardiac Monitoring.* Sensors, 2021. **21**(14): p. 4937.
86. Menvielle, L., A.-F. Audrain-Pontevia, and W. Menvielle, *The digitization of healthcare: new challenges and opportunities.* 2017: Springer.
87. Ponemon, I., *Sixth annual benchmark study on privacy & security of healthcare data.* 2016, Technical report.
88. Punter, T., et al. *Conducting on-line surveys in software engineering.* in *2003 International Symposium on Empirical Software Engineering, 2003. ISESE 2003. Proceedings.* 2003. IEEE.
89. Linaker, J., et al., *Guidelines for conducting surveys in software engineering v. 1.1.* Lund University, 2015.
90. Groves, R.M., et al., *Survey methodology.* 2011: John Wiley & Sons.
91. Garousi, V. and J. Zhi, *A survey of software testing practices in Canada.* Journal of Systems and Software, 2013. **86**(5): p. 1354-1376.
92. Garousi, V. and T. Varma, *A replicated survey of software testing practices in the Canadian province of Alberta: What has changed from 2004 to 2009?* Journal of Systems and Software, 2010. **83**(11): p. 2251-2262.
93. Saunders, M., P. Lewis, and A. Thornhill, *Research methods for business students.* 2009: Pearson education.
94. Henseler, J., C.M. Ringle, and R.R. Sinkovics, *The use of partial least squares path modeling in international marketing*, in *New challenges to international marketing.* 2009, Emerald Group Publishing Limited.
95. O'brien, R.M., *A caution regarding rules of thumb for variance inflation factors.* Quality & quantity, 2007. **41**(5): p. 673-690.

96. Fornell, C. and D.F. Larcker, *Evaluating structural equation models with unobservable variables and measurement error.* Journal of marketing research, 1981. **18**(1): p. 39-50.
97. Wong, K.K.-K., *Partial least squares structural equation modeling (PLS-SEM) techniques using SmartPLS.* Marketing Bulletin, 2013. **24**(1): p. 1-32.
98. Hair, J.F., C.M. Ringle, and M. Sarstedt, *PLS-SEM: Indeed a silver bullet.* Journal of Marketing theory and Practice, 2011. **19**(2): p. 139-152.
99. Dijkstra, T.K. and J. Henseler, *Consistent and asymptotically normal PLS estimators for linear structural equations.* Computational statistics & data analysis, 2015. **81**: p. 10-23.
100. Senapathi, M. and A. Srinivasan. *An empirical investigation of the factors affecting agile usage.* in *Proceedings of the 18th international conference on evaluation and assessment in software engineering.* 2014.
101. Fielding, N.G., N.F.R.M. Lee, and R.M. Lee, *Computer analysis and qualitative research.* 1998: Sage.
102. Runeson, P. and M. Höst, *Guidelines for conducting and reporting case study research in software engineering.* Empirical software engineering, 2009. **14**(2): p. 131-164.
103. Yin, R.K., *Case study research: Design and methods.* Vol. 5. 2009: sage.
104. Wohlin, C., et al., *Experimentation in software engineering.* 2012: Springer Science & Business Media.
105. Zhou, X., et al. *A map of threats to validity of systematic literature reviews in software engineering.* in *2016 23rd Asia-Pacific Software Engineering Conference (APSEC).* 2016. IEEE.
106. Khan, A.A., et al., *Systematic literature review and empirical investigation of barriers to process improvement in global software development: Client–vendor perspective.* Information and Software Technology, 2017. **87**: p. 180-205.
107. Akbar, M.A., et al., *Success factors influencing requirements change management process in global software development.* Journal of Computer Languages, 2019. **51**: p. 112-130.
108. Akbar, M.A., et al., *Requirements Change Management Challenges of Global Software Development: An Empirical Investigation.* IEEE Access, 2020. **8**: p. 203070-203085.
109. Rafi, S., et al., *Multicriteria based decision making of DevOps data quality assessment challenges using fuzzy TOPSIS.* IEEE Access, 2020. **8**: p. 46958-46980.
110. Akbar, M.A., et al., *Identification and prioritization of DevOps success factors using fuzzy-AHP approach.* Soft computing, 2020: p. 1-25.
111. Garvin, D.A., *Managing quality: The strategic and competitive edge.* 1988: Simon and Schuster.
112. Murdoch, Jonathan. "The spaces of actor-network theory." Geoforum 29.4 (1998): 357-374.